\documentclass[]{interact}

\usepackage{epstopdf}
\usepackage[caption=false]{subfig}

\usepackage[numbers,sort&compress,merge]{natbib}
\bibpunct[, ]{[}{]}{,}{n}{,}{,}
\renewcommand\bibfont{\fontsize{10}{12}\selectfont}

\usepackage{multirow}
\usepackage{tabularx}
\usepackage{graphicx}
\usepackage{xcolor}
\usepackage{float} 
\newcommand\T{\rule{0pt}{2.6ex}}       
\newcommand\B{\rule[-1.2ex]{0pt}{0pt}} 
\def\urlprefix{}
\def\url#1{}
\newcolumntype{Y}{>{\centering\arraybackslash}X}
\usepackage{dcolumn,booktabs}
\newcolumntype{d}[1]{D{.}{.}{#1}}
\newcommand\mc[1]{\multicolumn{1}{c}{#1}}
\newcommand\ml[1]{\multicolumn{1}{l}{#1}}

\theoremstyle{plain}

\theoremstyle{definition}

\theoremstyle{remark}

\begin{document}

\title{Born-Oppenheimer potentials for $\Pi$, $\Delta$, and $\Phi$ states of the hydrogen molecule} 

\author{
	\name{Micha{\l} Si{\l}kowski\thanks{michal.silkowski@fuw.edu.pl} and Krzysztof Pachucki}
	\affil{Faculty of Physics, University of Warsaw, Pasteura 5, 02-093 Warsaw, Poland} 
}

\maketitle

\begin{abstract}
	We report on accurate variational calculations of the Born-Oppenheimer potential for excited states of the hydrogen molecule with
	 $\Pi$, $\Delta$, and $\Phi$ symmetries. The obtained potential energy curves reach the relative precision of $10^{-9}$ or better 
	 along internuclear distances of 0.01 -- 20 au. Calculations rely on recursive evaluation of two-center two-electron
	molecular integrals with exponential functions in arbitrary precision arithmetics. 
	Our results for most of the states are the first ever reported, and for the previously calculated states
	constitute an improvement by several orders of magnitude.
\end{abstract}



\section{Introduction}
The hydrogen molecule is one of the most intensively studied two-electron system. Its rotational and vibrational energy levels in the ground electronic state are known with $10^{-9}$ relative accuracy,
by including nonadiabatic effects and relativistic and quantum electrodynamic corrections up to the order of $\alpha^6\,m$, where $\alpha\sim1/137$ is the fine-structure constant. 
Theoretical predictions agree very well the recent high-precision measurements
~\cite{Puchalski_2019,Liu_2009,Jennings_1986}, including those of the dissociation energy, which reach the accuracy of
$10^{-4}$ cm$^{-1}$ ~\cite{Cheng_2018,Altmann_2018,Hoelsch_2019}. Similar if not better experimental accuracy is achieved for the transitions to the excited electronic states,
but theoretical predictions are far less accurate here.

In our recent work we have performed calculations of Born-Oppenheimer (BO) potentials for excited states 
of all $n\,\Sigma^{+}$ symmetries with $n\le7$, for which we obtained the accuracy exceeding all the previously known results by 3-4 orders of magnitude.
In this work we extend those calculations to  $\Pi$, $\Delta$, and $\Phi$ excited states with the relative
accuracy of at least $10^{-9}$ for internuclear distances of 0.01 -- 20 au.
Some of these states have already been investigated in the literature. Namely, the most significant and accurate results at that time
were obtained primarily in a series of works by Ko\l os, Wolniewicz, and  Rychlewski~\cite{Koklos_1977,Ko_os_1982,Wolniewicz_1988,Rychlewski_1991,Wolniewicz_1995}, 
using the explicitly correlated exponential basis, called Ko\l os-Wolniewicz (KW) basis functions~\cite{Kolos_1966}.
Their results, however, disagreed with experimental values for $\Delta$ states. 
Jeziorski \emph{et al.} pointed out in Ref. \cite{Jeziorski_1997} that the angular factor for these states in Ref.
\cite{Wolniewicz_1995} was incomplete. This was verified and later corrected by Wolniewicz in Ref.  \cite{Wolniewicz2_1995}. 
It lowered BO curves by a few $\mathrm{cm}^{-1}$ for singlet and less than 1~cm$^{-1}$ for triplet $\Delta_{\mathrm{g}}$ states
and resolved discrepancies with the experimental data available at those times. 

Another method, based on explicitly correlated gaussian (ECG) functions, was employed in Ref. \cite{Komasa_1994, Cencek_1995} to obtain a potential curve for $C^1 \Pi_{\mathrm{u}}$ $(n=1)$.
The most recent contributions to the ab-initio Potential Energy Curves (PEC) of H$_2$ consist of a Free-Complement local Schr{\"o}dinger equation method
developed by Nakatsuji and collaborators \cite{Nakatsuji_2004,Nakatsuji_2015, Kurokawa_2018, Nakashima_2018, Kurokawa_2020} 
with potential curves for $\Sigma, \Pi, \Delta$, and $\Phi$ states. 
In this work we not only improve all these previous results by a few orders of magnitude, but also calculate many higher
excited states of $\Pi$, $\Delta$, and $\Phi$ symmetries and finally draw attention to possible further applications of Kolos-Wolniewicz basis.

\section{Method}

Our variational calculations utilize explicitly correlated exponential functions with polynomial dependence on interparticle
distances of the form~\cite{Kolos_1966},
\begin{eqnarray}
\Phi_{\{n\}} &=&
	e^{-y\,\eta_1 -x\,\eta_2 -u\,\xi_1 -w\,\xi_2}
	r_{12}^{n_0} \, \eta_1^{n_1} \, \eta_2^{n_2} \, \xi_1^{n_3} \, \xi_2^{n_4},
	\label{kw}
\end{eqnarray}
where $\eta_i$ and $\xi_i$ are proportional to confocal elliptic coordinates and are given by $\eta_i=r_{iA}-r_{iB}$,
$\xi_i=r_{iA}+r_{iB}$ with $i$ enumerating electrons and real $y,x,u,w$ nonlinear parameters subject to variational minimization.
By $\{n\}$ we denote an ordered set of interparticle coordinate exponents, $(n_0,n_1,n_2,n_3,n_4)$ which are
conventionally restricted by a shell parameter $\Omega$,
\begin{equation}
	\sum_{j=0}^4 n_j \le \Omega.
	\label{omega}
\end{equation}
If a symmetry restriction is imposed, the set of allowed $\{n\}$ is constrained even more for specific values of nonlinear parameters. 
By construction these functions depend on two-electron coordinates and account for the electronic
correlation via explicit dependence on the coordinate $r_{12}$. We emphasize that a construction of the basis according
to Eq. (\ref{omega}) allows for a very compact parametrization of the basis, because it is completely specified by just
four nonlinear parameters and an integer value of $\Omega$. To recapitulate, the trial wavefunction is represented as
\begin{equation}
	\Psi_{\Sigma^{+}} = \sum_{\{n\}} c_{\{n\}} \hat{S}^{\pm}_{AB} \hat{S}^{\pm}_{12} \Phi_{\{n\}},
	\label{wfsigma}
\end{equation}
where $\hat{S}^{\pm}_{AB} = 1 \pm P_{AB}$ and $P_{AB}$ permutes the nuclei $A$ and $B$, $\hat{S}^{\pm}_{12} = 1 \pm
P_{12}$ and $P_{12}$ interchange the two electrons, and appropriate $\pm$ signs are chosen
to fulfil the symmetry criteria for \textit{gerade}/\emph{ungerade} and \textit{singlet}/\textit{triplet} states. By
solving the secular equation one obtains linear coefficients $c_{\{n\}}$.
Such a form of wavefunction expansion is commonly referred to as the Ko{\l}os-Wolniewicz basis \cite{Kolos_1966}.
This nomenclature originated from the series of pioneering works of Ko{\l}os, Wolniewicz, and co-workers. It was introduced as a flexible generalization of the James-Coolidge basis
aiming to represent the electron density asymmetry between the two nuclei ubiquitous in excited states of H$_2$.

In previous calculations~\cite{Kolos_1966,Kolos_1976,Wolniewicz_1994,Wolniewicz_1995} involving tens or hundreds of KW functions in the basis, the set of $\{n\}$ was carefully
optimized by incremental selection of configurations that give the most significant contribution
to the energy. Here we resort to a much simpler rule, given by Eq.~(\ref{omega}), and introduce double basis functions
with common values of nonlinear parameters. For convenience we introduce a shorthand notation for our basis $S(\Omega_A,\Omega_B)$, 
where $S$ denotes a shorthand symmetry of the basis: either JC for generalized James-Coolidge ($x=y=0, u\neq w$)~\cite{James_1933} or KW for a general Kolos-Wolniewicz
basis with no additional restrictions. Values of $\Omega_A$ and $\Omega_B$ define the size of sectors, which are the
basis subsets carrying common independent nonlinear parameters and are constructed according to Eq. (\ref{omega}).

Construction of the basis by Eq. (\ref{omega}) for all the states considered in this work entails a universal and practical approach. 
It is expected, however, that more sophisticated selection of basis functions, such as subdivision into three $\Omega$,
each restricting the maximal value of exponents of $\eta_i$, $\xi_i$, and $r_{12}$ individually, as investigated by Sims~\cite{Sims_2006} in James-Coolidge basis or utilized in 
our approach to long-range exchange splitting~\cite{Silkowski_2020}, will lead to more compact wave function expansions.

\section{Angular factors}

Expansion of the type in Eq. (\ref{wfsigma}) is valid only for the states of $\Sigma^{+}$ symmetry. Extension of the
KW basis to states of higher angular momentum requires the introduction of angular factors.
The electronic state of a diatomic molecule is, among other symmetries, characterized\,by\,$\Lambda$ --\, an absolute value of the eigenvalue of the $\vec n \cdot \vec L$
operator, where $\vec n$ is a normalized vector parallel to the internuclear axis and $\vec L$ is the electronic angular momentum operator.
In order to enforce proper angular symmetry for $\Lambda \neq 0$ we construct symmetric, traceless $l$--th rank tensors,
\begin{eqnarray} \nonumber
	\chi^i_1 &=& \rho_1^i~\mathrm{for}~\Pi, \\  \nonumber
	\chi^{ij}_{11} &=& \rho_1^i \rho_1^j - \frac{1}{2} \delta_{\perp}^{ij}\, \rho_1^2~~ \mathrm{for}~\Delta, \\ 
	\chi^{ij}_{12} &=& \rho_1^{(i} \rho_2^{j)}  - \delta_{\perp}^{ij} \vec \rho_1 \cdot \vec \rho_2 ~~ \mathrm{for}~\Delta, \\  \nonumber
	\label{ang}
	\chi^{ijk}_{111} &=& \rho_1^i\rho_1^j\rho_1^k - \frac{1}{3}\rho_1^2\,\rho_1^{(k} \delta_{}^{ij)}  ~\mathrm{for}~\Phi, \\  \nonumber
	\chi^{ijk}_{112} &=& \rho_1^{(i}\rho_1^j\rho_2^{k)} - \frac{1}{3} \rho_1^2\,\rho_2^{(k}\,\delta^{ij)} - \frac{2}{3} \vec \rho_1 \cdot \vec \rho_2\,\rho_1^{(k}\,\delta^{ij)}~~\mathrm{for}~\Phi, \\  \nonumber
\end{eqnarray}
where $\rho^i = r^i-n^i\,\vec n\,\vec r$, $ \delta_{\perp}^{ij} = \delta^{ij}-n^i\,n^j$, $\vec n = \vec R/R$, and $(ijk)$ denotes symmetrization of indices. 
Such tensors represents irreducible representations of SO(2) rotations (around the internuclear axis) in the coordinate space.
Consequently, the total wavefunction for $\Pi$, $\Delta$, and $\Phi$ states, corresponding to
$\Lambda=1,2$, and 3, respectively, is expanded as
\begin{eqnarray} \nonumber
	\Phi^i_{\Pi} &=& \sum_{\{n\}}^{\Omega} c_{\{n\}} \hat{S}^{\pm}_{AB} \hat{S}^{\pm}_{12} \chi_1^i \Phi_{\{n\}}, \\ 
	\Phi^{ij}_{\Delta} &=& \sum_{\{n\}}^{\Omega} c_{\{n\}} \hat{S}^{\pm}_{AB} \hat{S}^{\pm}_{12} \chi^{ij}_{11}
	\Phi_{\{n\}} + \sum_{\{n\}}^{\Omega'} c'_{\{n\}} \hat{S}^{\pm}_{AB} \hat{S}^{\pm}_{12} \chi^{ij}_{12} \Phi'_{\{n\}}, \\ \nonumber
	\Phi^{ijk}_{\Phi} &=& \sum_{\{n\}}^{\Omega} c_{\{n\}} \hat{S}^{\pm}_{AB} \hat{S}^{\pm}_{12} \chi^{ijk}_{111}
	\Phi_{\{n\}} + \sum_{\{n\}}^{\Omega'} c'_{\{n\}} \hat{S}^{\pm}_{AB} \hat{S}^{\pm}_{12} \chi^{ijk}_{112} \Phi'_{\{n\}}.
	\label{wfang}
\end{eqnarray}

In the evaluation of matrix elements with the above functions, repeated Cartesian indices in bra and ket are summed
over, and the resulting expression is a
linear combination of $f$-integrals with various sets of $\{n\}$, which are defined as
\begin{equation}
f_{\{n\}}(R) = R\,\int \frac{d^3 r_1}{4\,\pi}\,\int \frac{d^3 r_2}{4\,\pi}\,
\frac{e^{ -w_1\,r_{12} - u\,\xi_1 - w\,\xi_2
                    - y\,\eta_1 - x\,\eta_2}}{r_{1A}\,r_{1B}\,r_{2A}\,r_{2B}}\,
r_{12}^{n_0-1}  \eta_1 ^{n_1} \eta_2 ^{n_2} \xi_1
^{n_3} \xi_2 ^{n_4}. \label{master}
\end{equation}
Efficient recursive evaluation of these integrals is the subject of our previous works~\cite{h2solv,rec_h2,kw}, and the details will not be repeated here.

The obtained expressions for matrix elements are too lengthy to be reported here, even those for the Hamiltonian with wavefunctions of the $\Pi$ symmetry.
Nevertheless, we have explicitly checked that our construction of angular factors yields the same linear combinations
of $f$-integrals, as with the use of either real or complex angular factors introduced by Jeziorski \emph{et
al.} in Ref.~\cite{Jeziorski_1997}.
In practical calculations, due to the length of these formulas, evaluation of matrix elements dominates the computation time for $\Delta$ and $\Phi$ symmetry, even at moderate basis sizes.


In pioneering calculations of the lowest gerade $\Delta$ states, namely $J$ and $S$ $^{1}\Delta_{\mathrm{g}}$ and $j$ and $s$
$^{3}\Delta_{\mathrm{g}}$, by Ko{\l}os and Rychlewski~\cite{Ko_os_1982, Rychlewski_1991} and later by Wolniewicz~\cite{Wolniewicz_1995,Wolniewicz2_1995},
the $\pi\pi$ terms were not included (the $\chi_{12}$ angular factor), as hinted by Jeziorski \emph{et al.} \cite{Jeziorski_1997}.
Its inclusion improves the adiabatic energies by a few cm${}^{-1}$ for $J^1\Delta_{\mathrm{g}}$ and $S^1\Delta_{\mathrm{g}}$ states and tenths
of cm${}^{-1}$ for triplet $j^3\Delta_{\mathrm{g}}$ and $s^3\Delta_{\mathrm{g}}$ states, as demonstrated in subsequent work by Wolniewicz~\cite{Wolniewicz2_1995}.
Together with the evaluation of nonadiabatic coupling by Yu and Dressler~\cite{Yu_1994}, the apparent discrepancy with experimental
values for dissociation energies of H$_2$ and D$_2$ ~\cite{Jungen_1990,Jungen_1992} has been resolved.




\begin{figure}[t]
\includegraphics[width=\columnwidth]{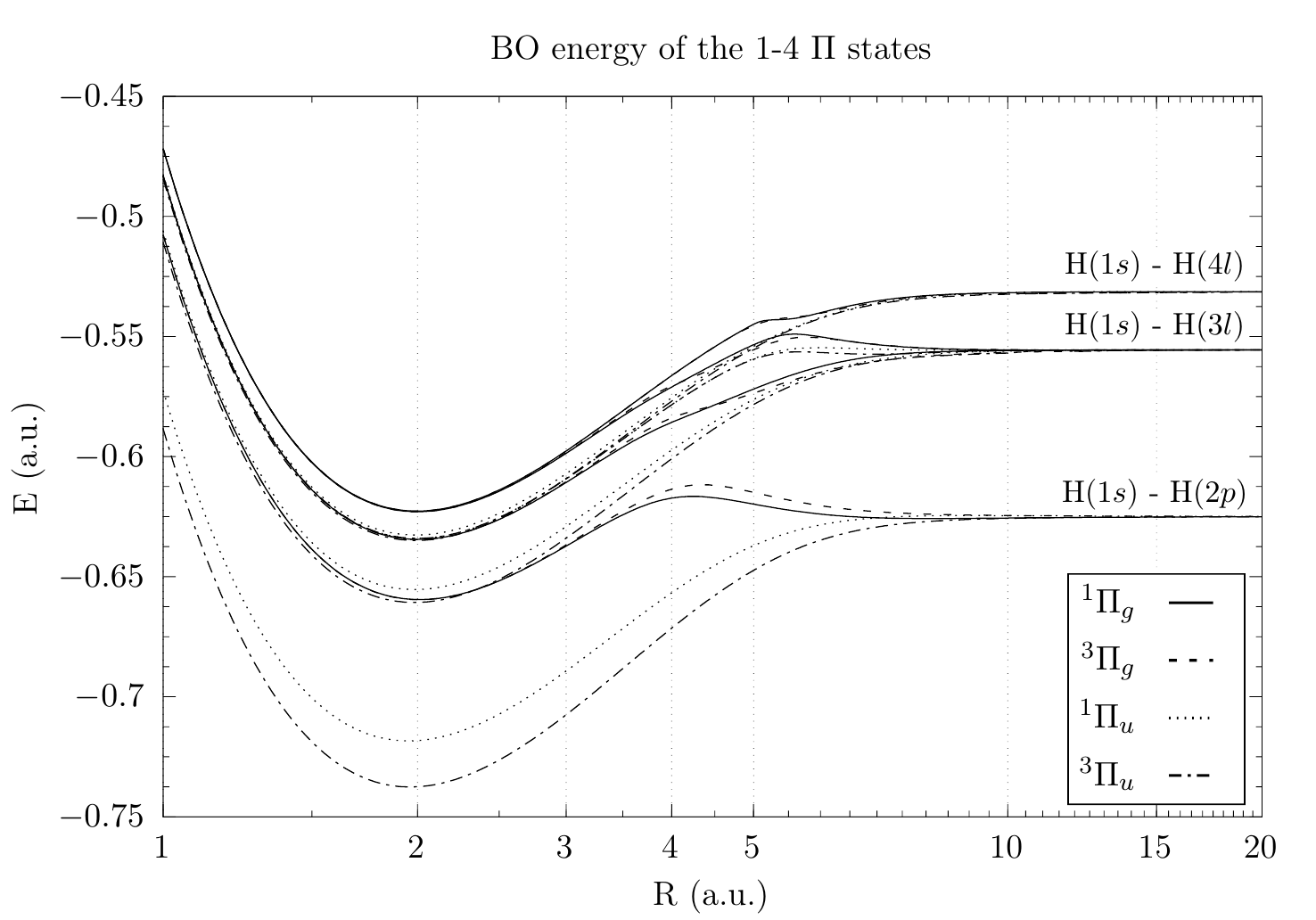}  
\caption{BO potential for $n=1..4$ $\Pi$ states.}
\label{pi}
\end{figure}

\section{Results}

We have calculated BO energies for $1-4\,\Pi$ and $1-3\,\Delta$ and $1-2\,\Phi$ states, and additionally
$3\,{}^1\Phi_{\mathrm{u}}$, and $3\,{}^3\Phi_{\mathrm{u}}$ states (38 in total) in 81 points of a nonuniform grid spanned within $R \in (0.01,20.0)$\,au.
In order to test the capability of the exponential basis, we have selected the $I^1\Pi_{\mathrm{g}}$, $i^3\Pi_{\mathrm{g}}$, $C^1\Pi_{\mathrm{u}}$, and
$c^3\Pi_{\mathrm{u}}$ states ($n=1$), for which we have ventured to exceptionally large bases consisting of two sectors with their
own nonlinear parameters:
JC(21,19) (54648 basis functions in total) in the range $R=0.01-6.0$\,au,
and KW(18,16) (53998 basis functions in total) in the range $R=6.5 - 20.0$\,au.
This allowed us to achieve 15 or more significant digits in the range $R=0.7 - 20\,au$, and can be regarded as the state-of-the-art of our current computational approach. 

In the remaining cases, calculations were performed with smaller bases, obtained by incrementing $\Omega$ of both
sectors by 1 until the extrapolated energies reached uncertainty no worse than $10^{-9}$ at all distances. This usually
required the use of the JC(14,12) basis in the range $0.01 - 6.0$ au and the KW(12,10) basis in $6.5 - 20.0$ au.
To reduce the computational cost of calculations with $\Delta$ and $\Phi$ symmetries, 
both sectors having different angular factor were set to carry the same nonlinear parameters. 
This significantly  reduced the number of $f$-integrals required for calculations of matrix elements.

All the potential curves have a well-pronounced single minimum around $R=2$~au and exhibit strong features of anticrossings 
and curve interactions in the region $R=4-8$~au, which are associated with the configuration mixing. 
Those interactions have a trend to become weaker with the increasing angular momentum of the excited electron.
Due to the high accuracy of the calculated potential, the dominating electronic configuration can
be deduced from comparison to energies of helium atom excited states and of an infinitely separated hydrogen H($1s$) -- H($nl$) atoms.
In particular, we observe, by reaching internuclear distances as small as $R=0.01$ au, that the energies smoothly evolve to
the corresponding value of the excited He atom~\cite{Drake2006}.
It follows from analysis of curves that each highly excited molecular state is dominated by a single atomic $1snl$ configuration at both
the united atom and the dissociation limit. The $nl$ configuration at both of those limits is usually not the same;
therefore, in such cases, when considering fixed $n$ in the molecular term (i.e., the $n$-th eigenvalue of the BO molecular
Hamiltonian), a character change of the excited electron has to occur at least once along the PEC.
Understanding of the electronic character of a molecule in terms of singly excited $1snl$ atomic configurations originates from the pioneering works on molecular binding theory
of Hund, Mulliken and many others~\cite{Mulliken_1928,Mulliken_1932,Mulliken_1955} and can be qualitatively described
with the correlation diagrams~\cite{Sharp_1971}. Detailed analysis of electronic characters along the PECs is out of scope of the current report
and in this work we mainly focus on the computational aspects and accuracy comparison with previous ab-initio results. For a meticulous analysis of electronic
configuration evolution as a function of $n$ and $R$, we refer to the extensive analysis of Corongiu and Clementi~\cite{Corongiu_2010}.

\subsection{States of $\Pi$ symmetry}

Although the basis  for the $\Pi$ state could in principle be built with a single sector, use of the second sector with different nonlinear parameters has
proven to be advantageous. Primarily, it clearly makes the basis more flexible with the introduction of another set of nonlinear parameters.
Secondarily, multi-sector bases are routinely utilized in atomic calculations with Hylleraas bases, where the largest sector
spans the most diffuse length scales with nonlinear parameters bounded from below by $\sqrt{-2\,m\,E}$, 
and subsequent sectors aim to improve the shorter scale range to the nucleus. 
Optimal values of parameters obtained in the present calculations for H$_2$ suggest that spanning more diffuse functions is
usually energetically more important than the short-range motion in the vicinity of the nuclei, and the introduction of a
second sector seems especially relevant in the regions without the domination of a single electronic configuration.

Our numerical results for $1-4~\Pi$ states are plotted in Fig. \ref{pi} and presented in the Supplementary Materials in Tables S1-S16.
In order to compare with the previous results of Refs. \cite{Kurokawa_2020,Datta_2010,Komasa_1994,Wolniewicz_1995} 
we select $R=2$ au and present value of potentials in Table \ref{tab:pi}. To test capabilities of our method, we have
also evaluated a few $\Pi_{\mathrm{u}}$ states with $n=5,6,10$ at a single point $R=2.0$\,au.

In Ref.~\cite{Datta_2010} an alternative approach to solving the Schr\"odinger equation in terms of stochastic
evaluation of the path integral with the
Schr\"odinger potential was introduced as an  energy estimator. In general these results were in good agreement with previous results. 
However, few $\sigma$ discrepancy was observed for the $I^1\Pi_{\mathrm{g}}$ state, for which their initial result of
-0.659\,515\,9(3) appeared to be discrepant by a few $\sigma$ with the variational result $-0.659\,515\,055\,5$ of
Wolniewicz \cite{Wolniewicz_1995}.
Curiously enough, by taking four times larger statistical sample, more specifically four times more integration paths in the
generalized Feynman-Kac method (GFK), they tightened the result to -0.659\,515\,54(6), thus suggesting a significant
improvement over the result of Wolniewicz.
Comparison of both results with ours reveals that result of Ref.~\cite{Datta_2010} overshoots the variational limit by almost
the same amount as the difference between our result and that of Wolniewicz. This indicates that the
convergence of the GFK method to the exact energy is much slower than anticipated and the extrapolation uncertainty was much too optimistic.

\begin{table}[H]
	\centering
	\footnotesize
	\caption{Comparison of BO energy near minimum for selected $\Pi$ states.}
\begin{tabular}{l d{1.3} d{3.20} c }
		\hline
		\hline
		\mc{method/basis size}		   &  \mc{$R/\mathrm{Bohr}$} & \mc{$E/\mathrm{hartree}$} & Ref. \T \\
		\hline
		\hline
		\multicolumn{4}{c}{I state~($n$=1)\,$\mathbf{1s2p^1\Pi_g}$} \T \\
		KW/75	&	2.0	&	-0.659\,511\,60	&	\cite{Koklos_1977} \\
		JC(5,3)/140	&	2.0	&	-0.659\,512\,139	&	this work \\
		KW/193	&	2.0	&	-0.659\,515\,055\,5	&	\cite{Wolniewicz_1995} \\
		JC(6,4)/272	&	2.0	&	-0.659\,515\,272	&	this work \\
		GFK	&	2.0	&	 -0.659\,515\,54(6)	&	\cite{Datta_2010} \\
		JC(21,19)/53636	&	2.0	&	-0.659\,515\,\underline{340\,754\,017}(6)	&			this work \\
			 \hline
			 \multicolumn{4}{c}{C state~($n$=1)\,$\mathbf{1s2p^1\Pi_u}$} \T \\
		KW/150	&	1.952	&	-0.718\,366\,655	&		\cite{Wolniewicz_1988} \\
		ECG/149	&	1.952	&	-0.718\,367\,778	&		\cite{Komasa_1994} \\
		JC(5,0.280,0.745,3,0.618,0.675)/168	&	1.952	&	-0.718\,367\,796\,214	&		this work \\
		KW/449	&	1.952	&	-0.718\,367\,979	&		\cite{Wolniewicz_1994} \\
		ECG/600	&	1.952	&	-0.718\,368\,027	&		\cite{Cencek_1995} \\
		JC(8,6)/917	&	1.952	&	-0.718\,368\,027\,687	&		this work \\
		JC(21,19)/53636	&	1.952	&	-0.718\,368\,0\underline{30\,147\,144\,23}(7)	&			this work \\
			 \hline
			 \multicolumn{4}{c}{D state~($n$=2)\,$\mathbf{1s3p^1\Pi_u}$} \T \\
		CI/50	&	2.0	&	-0.655\,035	&		\cite{Rothenberg_1966} \\
		KW/80	&	2.0	&	-0.655\,299\,82	&		\cite{Kolos_1976} \\
		KW/150	&	2.0	&	-0.655\,325\,841	&		\cite{Wolniewicz_1988} \\
		FC-LSE\textsuperscript{a}/1050	&	2.0	&	-0.655\,328\,191	&		\cite{Kurokawa_2020} \\
		JC(6,4)/316	&	2.0	&	-0.655\,328\,203	&		this work \\
		KW/724	&	2.0	&	-0.655\,328\,261	&			\cite{Wolniewicz_2003} \\
		JC(7,5)/552	&	2.0	&	-0.655\,328\,266\,813	&		this work \\
		JC(8,6)/917	&	2.0	&	-0.655\,328\,274\,470	&		this work \\
		JC(14,12)/9080	&	2.0	&	-0.655\,328\,2\underline{76\,407}(7)	&			this work \\
			 \hline
			 \multicolumn{4}{c}{V state~($n$=3)\,$\mathbf{1s4f^1\Pi_u}$} \T \\
		KW	&	2.0	&	-0.633\,939	&		\cite{Rothenberg_1966} \\
		MRCI/173	&	2.0	&	-0.632\,582\,84	&		\cite{Drira_1999} \\
		MRCI/284	&	2.0	&	-0.634\,045\,91	&		\cite{Spielfiedel_2003} \\
		FC-LSE\textsuperscript{a}/1050	&	2.0	&	-0.634\,036\,945	&		\cite{Kurokawa_2020} \\
		KW/724	&	2.0	&	-0.634\,058\,929	&			\cite{Wolniewicz_2003} \\
		JC(8,6)/917	&	2.0	&	-0.634\,059\,096\,079	&		this work \\
		JC(14,12)/9080	&	2.0	&	-0.634\,05\underline{9\,104\,699\,22}(8)	&			this work \\
			 \hline
			 \multicolumn{4}{c}{D' state~($n$=4)\,$\mathbf{1s4p^1\Pi_u}$} \T \\
		KW &	2.0	&	-0.632\,391	&		\cite{Rothenberg_1966} \\
		MRCI/284	&	2.0	&	-0.632\,606\,24	&		\cite{Spielfiedel_2003} \\
		FC-LSE\textsuperscript{a}/1050	&	2.0	&	-0.632\,669\,650	&		\cite{Kurokawa_2020} \\
		KW/724	&	2.0	&	-0.632\,670\,21	&			\cite{Wolniewicz_2003} \\
		JC(8,6)/917	&	2.0	&	-0.632\,670\,221\,022	&		this work \\
		JC(14,12)/9080	&	2.0	&	-0.632\,670\,2\underline{22\,456}(6)	&			this work \\
			 \hline
			 \multicolumn{4}{c}{($n$=5)\,$\mathbf{1s5f^1\Pi_u}$} \T \\
		FC-LSE\textsuperscript{a}/1050	&	2.0	&	-0.621\,700\,378	&		\cite{Kurokawa_2020} \\
		JC(8,6)/917	&	2.0	&	-0.622\,727\,031	&		this work \\
		KW(12,10)/9191	&	2.0	&	-0.62\underline{2\,727\,143\,343}(3)	&			this work \\
			 \hline
			 \multicolumn{4}{c}{($n$=6)\,$\mathbf{1s5p^1\Pi_u}$} \T \\
		FC-LSE\textsuperscript{a}/1050	&	2.0	&	-0.620\,612\,4	&		\cite{Kurokawa_2020} \\
		JC(8,6)/917	&	2.0	&	-0.622\,010\,993	&		this work \\
		JC(14,12)/9080	&	2.0	&	-0.62\underline{2\,011\,008\,773}(5)&			this work \\
			 \hline
			 \multicolumn{4}{c}{($n$=10)\,$\mathbf{1s7f^1\Pi_u}$} \T \\
		Full-CI/320	&	2.0	&	-0.603\,59	&		\cite{Corongiu_2010} \\
		KW(8,6)/1749	&	2.0	&	-0.612\,873\,187\,073	&		this work \\
		KW(14,12)/17816	&	2.0	&	-0.6\underline{12\,873\,232\,0}(1)	&			this work \\
			 \hline
			 \multicolumn{4}{c}{($n$=10)\,$\mathbf{1s7p^3\Pi_u}$} \T \\
		Full-CI/320	&	2.0	&	-0.605\,11	&		\cite{Corongiu_2010} \\
		KW(8,6)/1749	&	2.0	&	-0.613\,021\,263\,099	&		this work \\
		KW(14,12)/17816	&	2.0	&	-0.6\underline{13\,021\,373\,965}(5)	&			this work \\
			 \bottomrule
\end{tabular}
\begin{flushleft}
\tabnote{\textsuperscript{a}Free Complement local Schr{\"o}dinger equation method}
\end{flushleft}

\label{tab:pi}

\end{table}

\begin{figure}[H]
	\centering
	\subfloat[$(n=1)$~$2p\pi^1\Pi_{\mathrm{u}}$ (C state)]{
	\resizebox*{0.48\textwidth}{!}{\includegraphics[width=.48\linewidth]{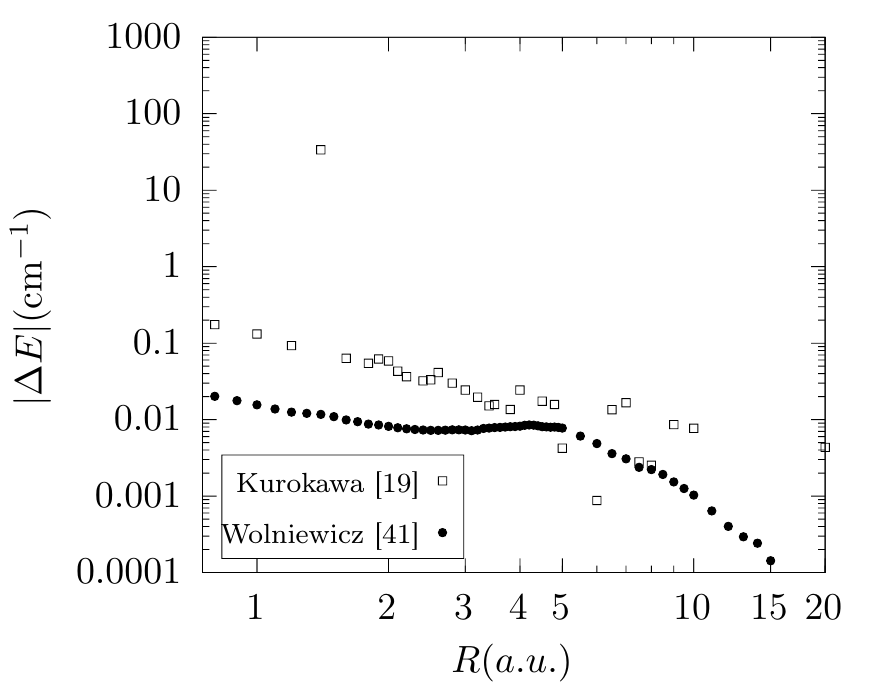}}}
	\subfloat[$(n=3)$~$4f\pi^1\Pi_{\mathrm{u}}$ (V state)]{
	\resizebox*{0.48\textwidth}{!}{\includegraphics[width=.48\linewidth]{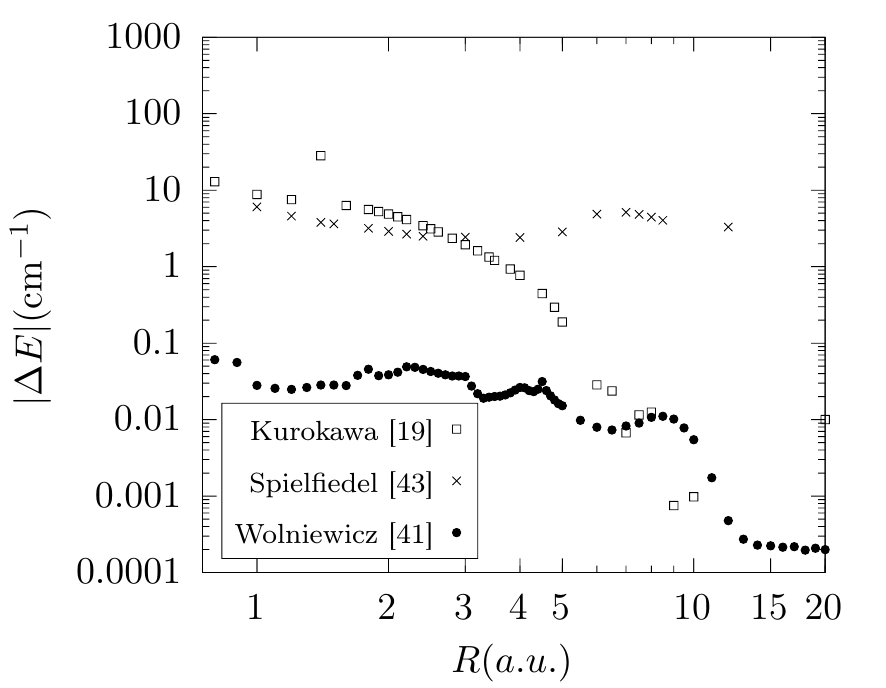}}}
	\\
	\subfloat[$(n=3)$~$5d\pi^3\Pi_{\mathrm{g}}$ (w state)]{
	\resizebox*{0.48\textwidth}{!}{\includegraphics[width=.48\linewidth]{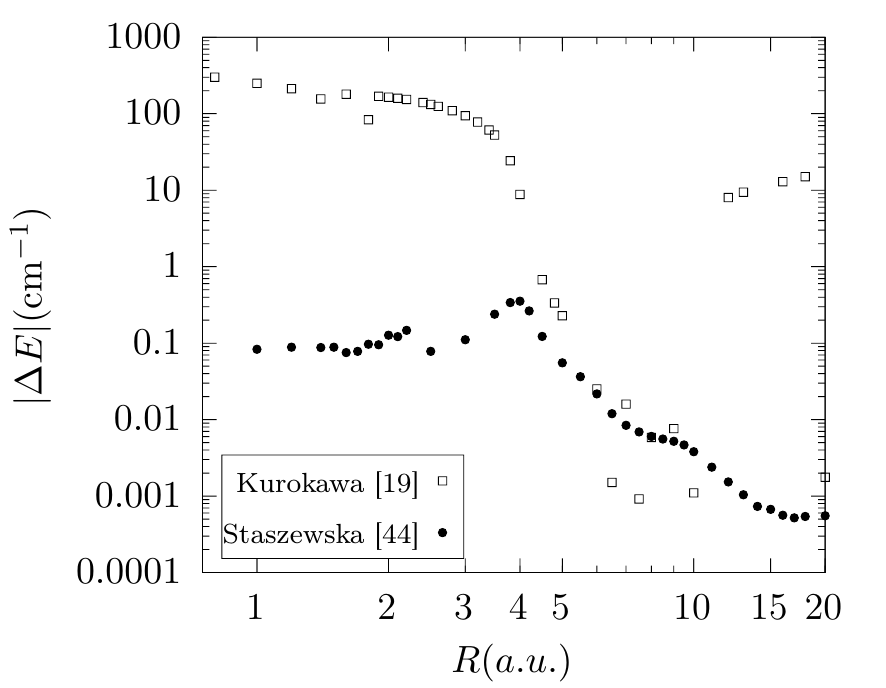}}}
	\subfloat[$(n=2)$~$4d\delta^1\Delta_{\mathrm{g}}$ (S state)]{
	\resizebox*{0.48\textwidth}{!}{\includegraphics[width=.48\linewidth]{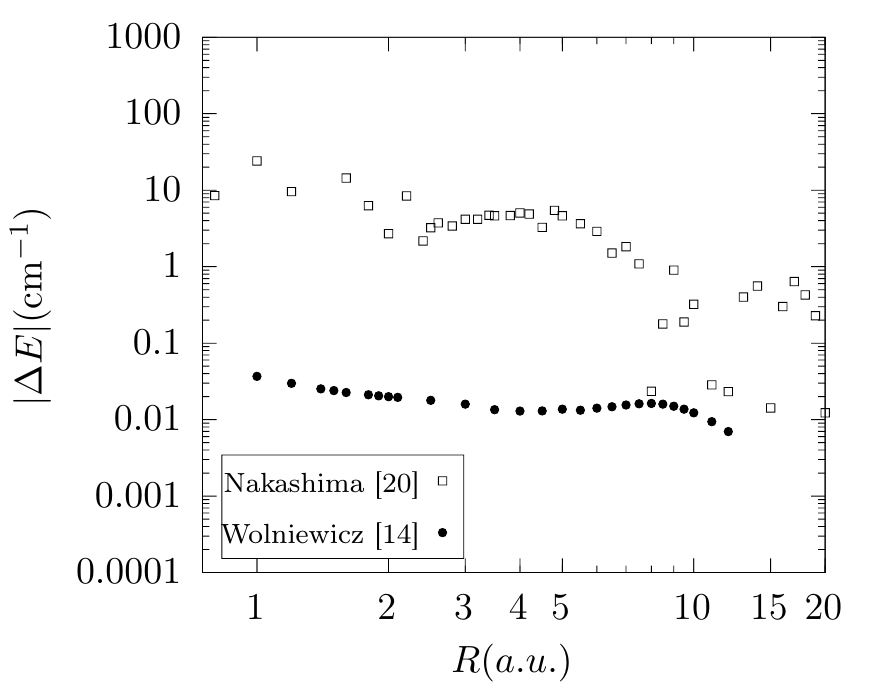}}}
	\caption{Comparison of BO energies for selected $\Pi$ and $\Delta$ states as a function of $R$.}
	\label{comp}
\end{figure}

In Fig. \ref{comp}, we have plotted the difference between our results and that of Refs.
\cite{Wolniewicz_2003,Staszewska_2001,Spielfiedel_2003,Kurokawa_2020,Nakashima_2018} as a function of $R$ for a few selected $\Pi$ states differing in the singlet/triplet and gerade/ungerade symmetry, as well as in the character of the excited electron.
Considering the energy differences between our results and those of
Refs.~\cite{Kurokawa_2020,Komasa_1994,Wolniewicz_1995} for various $\Pi$ states as a function of $R$,
we conclude that rather old results of Wolniewicz \cite{Wolniewicz_2003} for $1-4^1\Pi_{\mathrm{u}}$ states are accurate to at least 0.1 cm$^{-1}$.
Interestingly, the agreement with Ref. \cite{Kurokawa_2020} within $\sim$ 1 cm$^{-1}$ accuracy is observed for the
$1,2~\Pi$ states of $p$ and $d$ character. In contrary the discrepancy is much bigger for the states dominated by the
configurations with higher $l$, where the discrepancy reaches even hundreds of cm$^{-1}$, see Fig \ref{comp}c.
Most likely it is due to the poor representation of singly-excited configurations with large values of $l$, especially
of $f,g$, and $h$ character, and could be associated with the absence of terms proportional to H(1s)\,--\,H($n=4$) in the initial function of their method.
This absence is already noted by the Authors of Ref.~\cite{Kurokawa_2020}.

\subsection{States of $\Delta$ symmetry}

\begin{table}[H]
	\centering
	\footnotesize
	\caption{Comparison of BO energy $\Delta$ states at $R=2.0\,au$. To the Authors' best knowledge, these are all the ab-initio results of $\Delta$ symmetry available in the literature. $\delta$ is the difference with respect to the best available result.}
\begin{tabular}{l d{1.3} d{3.20} d{3.3} c }
		\hline
		\hline
		\mc{method \& basis size}		   &  \mc{$R/\mathrm{Bohr}$} & \mc{$E/\mathrm{hartree}$} & \mc{$\delta /\mathrm{cm^{-1}}$} \T \\
		\hline
		\hline
	\multicolumn{5}{c}{J state~($n$=1)~$\mathbf{1s3d^1\Delta_g}$} \T \\
		Full-CI/320	&	2.0	&	-0.657\,53	&	10.57	&	\cite{Corongiu_2010} \\
		FC-LSE	&	2.0	&	-0.657\,565(52)	&	2.89		&	\cite{Nakashima_2018} \\
		VMC	&	2.0	&	 -0.657\,577\,61(4)	&	0.12		&	\cite{Datta_2010} \\
		GFK	&	2.0	&	 -0.657\,578\,0(3)	&	0.04		&	\cite{Datta_2010} \\
		KW/394	&	2.0	&	-0.657\,578\,084\,4	&	0.02	&	\cite{Wolniewicz2_1995} \\
		JC(6,4)/316	&	2.0	&	-0.657\,578\,102\,4	&		0.016	&	this work \\
		JC(16,14)/16233	&	2.0	&	-0.657\,578\,\underline{175\,516}(2)	&		0.0	&	this work \\
			 \hline
		 \multicolumn{5}{c}{j state~($n$=1)~$\mathbf{1s3d^3\Delta_g}$} \T \\
		Full-CI/320	&	2.0	&	-0.657\,59	&	10.27	&	\cite{Corongiu_2010} \\
		FC-LSE	&	2.0	&	-0.657\,658(29)	&	-4.65	&	\cite{Nakashima_2018} \\
		KW/394	&	2.0	&	-0.657\,636\,708\,8	&	0.02	&	\cite{Wolniewicz2_1995} \\
		JC(6,4)/316	&	2.0	&	-0.657\,636\,722\,7	&		0.016	&	this work \\
		JC(16,14)/9080	&	2.0	&	-0.657\,636\,7\underline{96\,522}(2)	&		0.0	&	this work \\
			 \hline
			 \multicolumn{5}{c}{S state~($n$=2)~$\mathbf{1s4d^1\Delta_g}$} \T \\
		FC-LSE	&	2.0	&	-0.633\,651(40)	&	-2.65	&	\cite{Nakashima_2018} \\
		KW/70	&	2.0	&	-0.633\,607\,617	&	6.87	&	\cite{Rychlewski_1991} \\
		KW/394	&	2.0	&	-0.633\,638\,424\,0	&	0.11	&	\cite{Wolniewicz2_1995} \\
		JC(8,6)/917	&	2.0	&	-0.633\,638\,871	&		0.012	&	this work \\
		JC(14,12)/16233	&	2.0	&	-0.633\,638\,\underline{926\,5}(2)	&		0.0	&	this work \\
			 \hline
			 \multicolumn{5}{c}{s state~($n$=2)~$\mathbf{1s4d^3\Delta_g}$} \T \\
		FC-LSE	&	2.0	&	-0.633\,697(25)	&	-5.72	&	\cite{Nakashima_2018} \\
		KW/70	&	2.0	&	-0.633\,657\,867	&	2.87	&	\cite{Rychlewski_1991} \\
		KW/394	&	2.0	&	-0.633\,670\,563\,6	&	0.08	&	\cite{Wolniewicz2_1995} \\
		JC(14,12)/9080	&	2.0	&	-0.633\,670\,\underline{925\,50}(8)	&		0.0	&	this work \\
			 \hline
			 \multicolumn{5}{c}{($n$=1)~$\mathbf{1s4f^1\Delta_u}$} \T \\
		Full-CI/320	&	2.0	&	-0.633\,93	&	0.91	&	\cite{Corongiu_2010} \\
		MRCI\textsuperscript{b}/284	&	2.0	&	-0.633\,923\,21	&	2.40	&	\cite{Spielfiedel_2003} \\
		JC(6,4)/272	&	2.0	&	-0.633\,934\,070	&		0.016	&	this work \\
		JC(14,12)/9080	&	2.0	&	-0.633\,93\underline{4\,144\,964}(1)	&		0.0	&	this work \\
			 \bottomrule
\end{tabular}
\begin{flushleft}
\tabnote{
	\textsuperscript{a} Free Complement local Schr{\"o}dinger equation method \\ 
	\textsuperscript{b} Multi-reference Configuration Interaction
}
\end{flushleft}
\label{tab:delta}

\end{table}

\begin{figure}[h]
\includegraphics[width=\columnwidth]{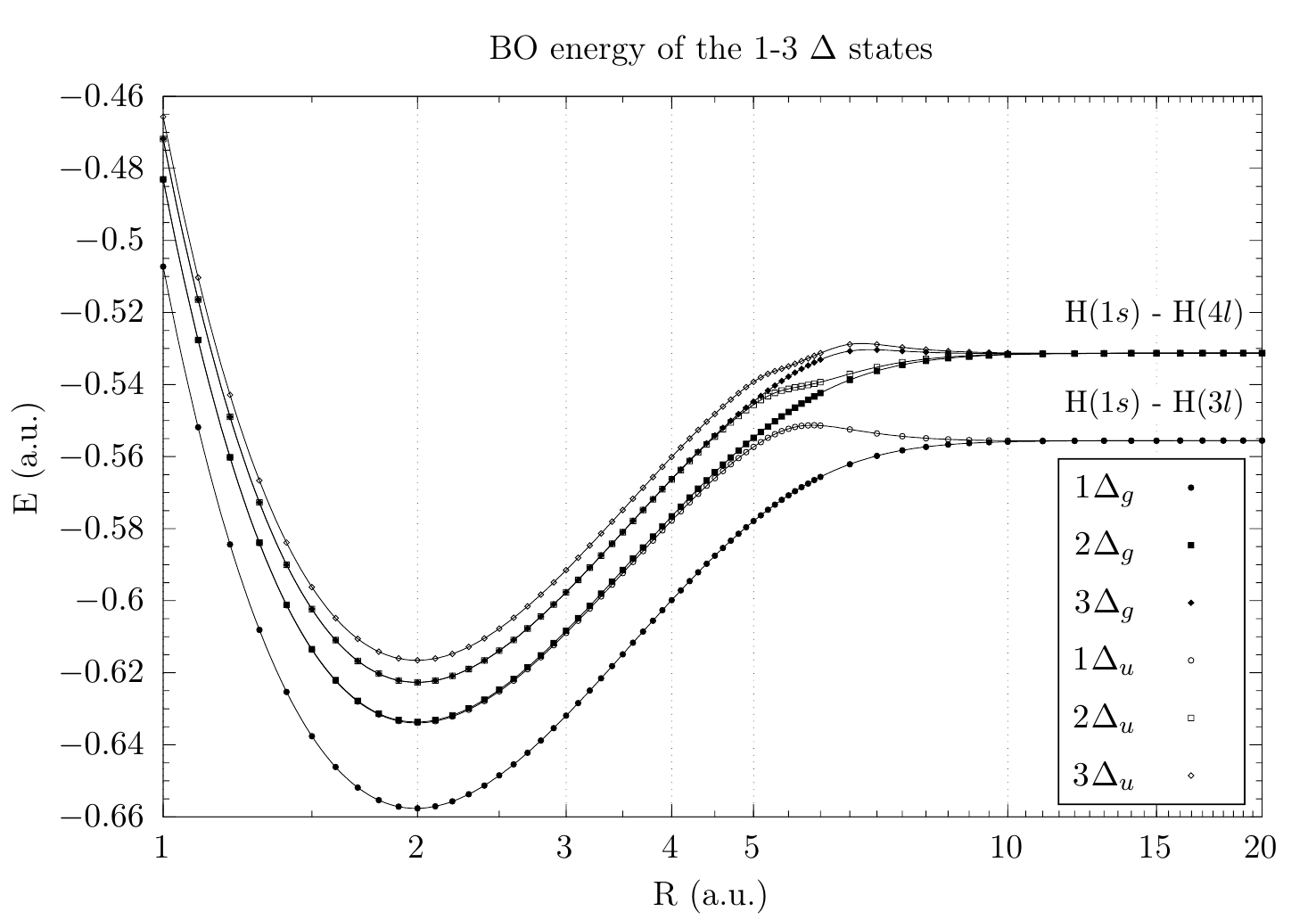}  
\caption{BO potential for $n=1..3$ $\Delta$ states. The difference between singlet and triplet states for specific
	gerade/ungerade symmetry and fixed $n$ is indistinguishable on this scale. It amounts to a few cm\textsuperscript{-1} at all
distances, peaking at around $R=3$ au, where it reaches $\sim$ 15\,cm\textsuperscript{-1} for $1\,\Delta_{\mathrm{g}}$.}
\label{delta}
\end{figure}

Our numerical results for $1-3\, \Delta$ states are presented in the Supplementary Material in Tables S17-S28, and are shown in Fig. \ref{delta}.
Moreover, in Table~\ref{tab:delta}. we compare our results for $\Delta$ symmetry at $R=2.0$\,au (vicinity of the minimum) with all
the ab-initio results available in the literature. Comparison reveals that Wolniewicz's results for $J$ and $S$
$^{1}\Delta_{\mathrm{g}}$ and $j$ and $s$ $^{3}\Delta_{\mathrm{g}}$ states are accurate up to $0.1$ cm$^{-1}$, which is roughly the uncertainty
he had originally assigned to his calculations~\cite{Wolniewicz2_1995}. This Table reveals that for the $S$ state, the basis JC(14,12) even with
different parameters for both sectors, gives rather slow convergence to the CBS limit; therefore, the JC basis is far from optimal for this state.
By virtue of high accuracy of our curves we can unambigously assign state configurations by direct comparison with
helium values~\cite{Drake2006}. In particular $1,2,3\,\Delta_{\mathrm{g}}$ and $1,2,3\,\Delta_{\mathrm{u}}$ states start at $R=0$ as $1s3d$,
$1s4d$, $1s5d$, $1s4f$, $1s5f$, and $1s6f$, respectively and this configurations propagate at least up to $R=5$~au, with
the exception of $3\,\Delta_{\mathrm{g}}$ states which sharply switch to $1s5g$ configuration at $R \sim 0.5$ au.

\begin{figure}[h]
\includegraphics[width=\columnwidth]{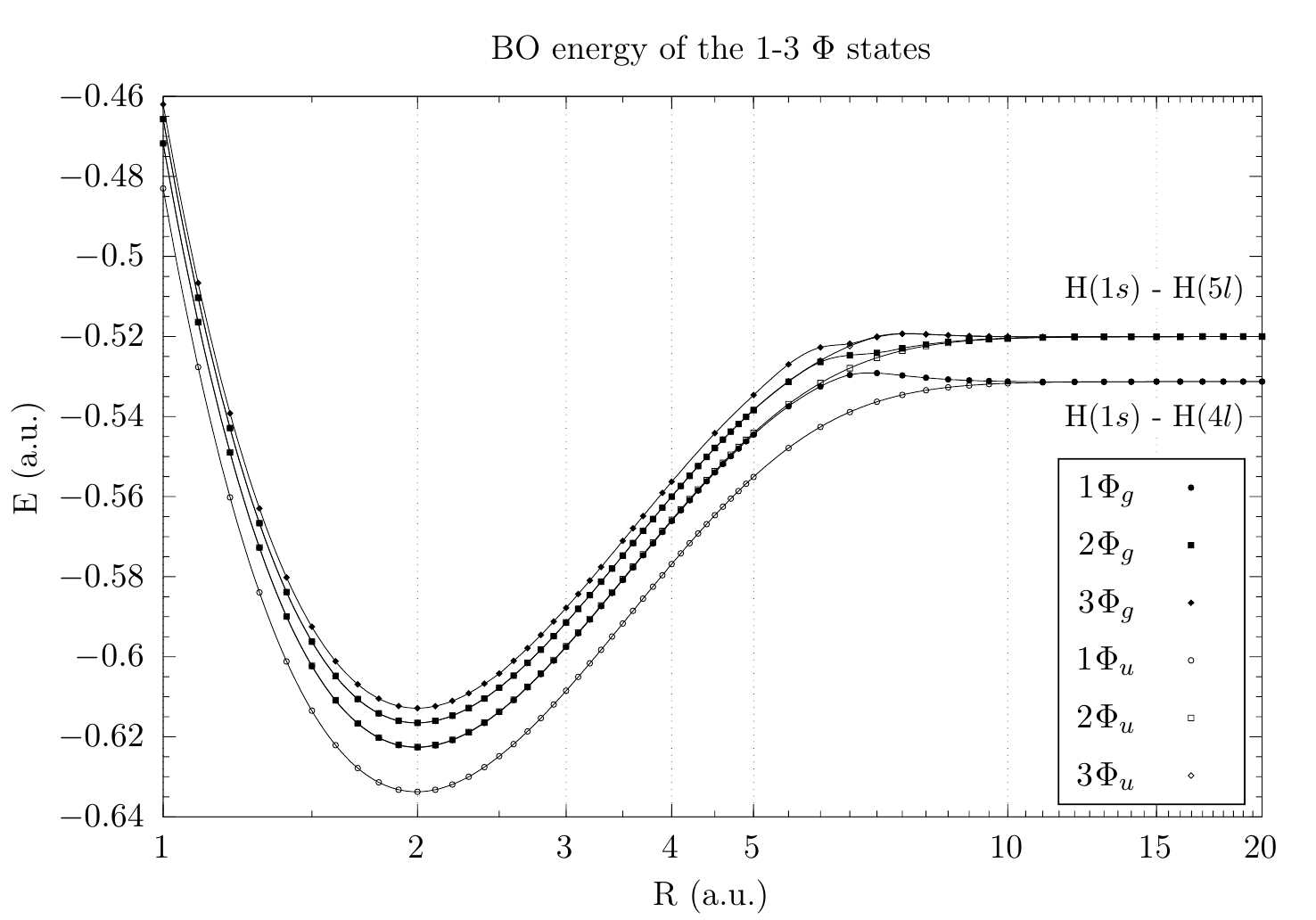}  
\caption{BO potential for $n=1..3$ $\Phi$ states. The difference between singlet and triplet states for specific gerade/ungerade symmetry and fixed $n$ is indistinguishable on this scale.}
\label{phi}
\end{figure}

\subsection{States of $\Phi$ symmetry}
\begin{table}[H]
	\centering
	\footnotesize
	\caption{Comparison of BO energy of $\Phi$ states at $R=2.0\,au$ with, to the Authors' best knowledge, the only ab-initio
results available in the literature for this symmetry. Underlined digits present an improvement with respect to the previous best value.}

{\renewcommand{\arraystretch}{1.1}
\begin{tabular}{l d{3.20} d{3.20} }
		\hline
		\hline
		\mc{method \& basis size}	 &  \ml{$E/\mathrm{hartree}$} & \ml{$E/\mathrm{hartree}$}  \T \\
		\hline
		\T
							                   & \multicolumn{1}{l}{$(n=1)~\mathbf{1s4f^1\Phi_u}$} & \multicolumn{1}{l}{$(n=1)~\mathbf{1s4f^3\Phi_u}$} \\
Full-CI/320, Ref. \cite{Corongiu_2010}	                   & -0.633\,72 &			-0.633\,72					\\	
FC-LSE\textsuperscript{a}, Ref. \cite{Nakashima_2018}		 & -0.633\,746(24) &			-0.633\,746(24)					\\
JC\textsuperscript{*}(5,3)/140	                         & -0.633\,733\,620 &			-0.633\,733\,849					\\
JC\textsuperscript{*}(15,13)/12240\textsuperscript{b}	                         & -0.6337\underline{33701792}(6) & -0.633\,7\underline{33\,908\,758}(2) \\
 \hline
							                   & \multicolumn{1}{l}{$(n=1)~\mathbf{1s4g^1\Phi_g}$} &
										 \multicolumn{1}{l}{$(n=1)~\mathbf{1s4g^3\Phi_g}$} \T\\
Full-CI/320, Ref. \cite{Corongiu_2010}                       &  			-0.622\,61								      &  			-0.622\,61							            	 \\
FC-LSE\textsuperscript{a}, Ref. \cite{Nakashima_2018}		 &  			-0.622\,53(17)				                        &  			-0.622\,53(17)				                        	 \\
JC\textsuperscript{*}(5,3)/140	                         &  			-0.622\,625\,783							      &  			-0.622\,625\,783						            	 \\
JC\textsuperscript{*}(15,13)/11832\textsuperscript{c}	                         & -0.622\,6\underline{28\,438\,589}(3) & -0.622\,6\underline{28\,439\,218}(3) \\
 \hline
							&	\multicolumn{1}{l}{$(n=3)~\mathbf{1s6h^1\Phi_u}$} & \multicolumn{1}{l}{$(n=3)~\mathbf{1s6h^3\Phi_u}$} \T \\
 Full-CI/320, Ref. \cite{Corongiu_2010}	&        		-0.613\,77								&-0.613\,77		\\
 JC\textsuperscript{*}(5,3)/140		&        		-0.616\,525\,695\,795						 &-0.616\,525\,695\,796\\
 JC\textsuperscript{*}(15,13)/12240\textsuperscript{d}		&			-0.61\underline{6\,525\,704\,654\,9}(2) &-0.61\underline{6\,525\,704\,656\,4}(2) \B \\
 \hline
\end{tabular}
\begin{flushleft}
\tabnote{
	\textsuperscript{*} the same nonlinear parameters in both basis sectors \\
	\textsuperscript{a} Free Complement local Schr{\"o}dinger equation method\\
	\textsuperscript{b} $u=0.173$,\,$w=0.760$\\
	\textsuperscript{c} $u=0.105$,\,$w=0.803$\\
	\textsuperscript{d} $u=0.082$,\,$w=0.760$
}
\end{flushleft}
}
\label{tab:phi}

\end{table}
Our numerical results for $1-2\, \Phi$, and $3\,\Phi_{\mathrm{u}}$ states are presented in the Supplementary Material in Tables S29-S38, and are shown in Fig. \ref{phi}.
In the literature $\Phi$ states were calculated only in Ref.~\cite{Nakashima_2018} using the Free Complement local Schr\"odinger equation (FC-LSE) method of Nakashima and
Nakatsuji~\cite{Nakatsuji_2015} and in Ref.~\cite{Corongiu_2010} using the Full-CI method. In this latter work the results
were presented only for a few selected states and at specific points corresponding to the energy minimum and dissociation limit, $R=2$ au and
$R=100$ au, respectively. Consequently, in Table~\ref{tab:phi} we present a comparison to those results at the curve minimum $R=2$ au.
It appears, that for the $\Phi$ state, FC-LSE functions do not span the complete Hilbert space with their choice of initial spatial functions. 
The deviation is slightly smaller than 1\,cm$^{-1}$, and thus larger than the uncertainty estimates.
Ultimately though, direct comparison with experimental data is cumbersome, because for such highly excited states the Rydberg electron
decouples from the nuclear axis due to nonadiabatic effects and $\Lambda$ is no longer a good quantum number.

Due to high angular momentum, the curves are very regular. The difference between neighbouring states propagates almost
unchanged from those of helium values~\cite{Drake2006} at $R=0$ up to around $R=5$ au. The singly-excited characters of $1,2\,\Phi_{\mathrm{g}}$ and $1,2,3\,\Delta_{\mathrm{u}}$ states start at
$R=0$ as $1s5g$, $1s6g$, $1s4f$, $1s5f$, and $1s6f$, respectively and this configurations propagate at least up to $R=5$~au, with
the exception of $3\,\Phi_{\mathrm{u}}$ states which sharply switch to $1s6h$ configuration at $R \sim 0.3$ au.

\section{Summary and Conclusions}

In order to compare our results with the plethora of transitions measured with accuracy reaching $\sim
0.001\,\mathrm{cm}^{-1}$, theoretical values of relativistic, QED, adiabatic, and nonadiabatic corrections are necessary.
The importance of the latter is arguably the most significant due to strong nonadiabatic couplings, because in contrast to the well-isolated $X\,{}^1\Sigma^{+}_{\mathrm{g}}$ ground state, energy differences of BO energies between neighboring
excited states can be as small as a few cm$^{-1}$. 
Even though some of the states with high $\Lambda$ are hardly accessible with spectroscopic
methods, knowledge about them has proven to be beneficial for extrapolation of high-$n$ molecular Rydberg states.
Incorporation of clamped-nuclei ab-initio potentials for relatively low $n$ has been found fruitful in approaches based on Multichannel Quantum Defect Theory (MQDT) for Rydberg states of H$_2$~\cite{Herzberg_1972,Sprecher_2014,Sprecher2_2014,Beyer_2018,Hoelsch_2018,Beyer_2019}.


In recent years explicitly correlated Gaussian (ECG) functions have achieved a great success in high-precision calculations of
hydrogen molecule isotopologues, ranging in applications from BO energy~\cite{Rychlewski_1994,Komasa_1994} and nonadiabatic
corrections~\cite{Puchalski_2018,Komasa_2019} to relativistic~\cite{Puchalski_2017,Czachorowski_2018} and QED~\cite{Piszczatowski_2009,Puchalski_2016} corrections.
All those calculations, however, were limited to the electronic $X^1\Sigma^{+}_{\mathrm{g}}$ ground state.

In the present work we have demonstrated that the Ko{\l}os-Wolniewicz bases still find a niche in
high-accuracy variational calculations of excited states of a diatomic two-electron molecule, especially for the states with high angular momentum.
Alongside with our previous calculations of $\Sigma^{+}$ states~\cite{Si_kowski_2021}, present calculations reconcile
with all the previous \emph{ab-initio} calculations of BO potentials of H$_2$ by elucidating the actual accuracy of the former calculations. 
Numerical uncertainty of Born-Oppenheimer potentials has been alleviated from a level of a few to the millionth parts of $\mathrm{cm}^{-1}$; hence, we have laid the foundation for the evaluation of further corrections to the energy
levels, lack of knowledge of which currently hinders comparison with state-of-the-art spectroscopic results.

\section*{Disclosure statement}
We declare no conflict of interest.

\section*{Acknowledgements}
The authors acknowledge support from the National Science Center (Poland) under Grant No. 2017/27/B/ST2/02459, and
M.S. acknowledges additional support under Grants No. 2020/36/T/ST2/00605 (Doctoral scholarship ETIUDA).

\bibliographystyle{tfo}

\end{document}


\section*{Supplementary Material}
\title{Supplementary material to {\em Born-Oppenheimer potentials for $\mathbf\Pi$, $\mathbf\Delta$, and $\mathbf\Phi$ states of the hydrogen molecule}}
\author{Micha\l\ Si{\l}kowski\footnote{michal.silkowski@fuw.edu.pl} and Krzysztof Pachucki}
\maketitle
{\centering
\textit{Faculty of Physics, University of Warsaw, Pasteura 5, 02-093 Warsaw, Poland} \\
({\small Version 0.1} dated: \today) \\
}
\vspace*{2.5em}

Contents: \\
Tables S1 to S38 are 5 column wide, where the columns contain (in order): $R$, $E$, $\langle V \rangle$, $\langle \nabla_1 \cdot \nabla_2 \rangle$, $\mathrm{d}E/\mathrm{d}R$.
All the quantities are given in atomic units. Each state is presented in a separate table, \\
Tables S1 to S16:  $1-4\,\Pi$ states,\\
Tables S17 to S28:  $1-3\,\Delta$ states,\\
Tables S29 to S38:  $1-2\,\Phi$, $3\,{}^1\Phi_\mathrm{u}$, and $3\,{}^3\Phi_\mathrm{u}$ states.\\

\input{../tables.tex}

